\documentclass{aa}
\usepackage{graphicx}
\usepackage{natbib}
\bibpunct{(}{)}{;}{a}{}{,}

\begin{document}

\hyphenation{drift-rate}

\title{Probing drifting and nulling mechanisms through their
interaction in PSR~B0809+74}
\titlerunning{Probing drifting and nulling mechanisms in PSR~B0809+74}

\author{
A.G.J. van Leeuwen       \inst{1}
\and 
B.W. Stappers            \inst{2,3}
\and
R. Ramachandran          \inst{2,3}
\and
J. M. Rankin             \inst{3,4}
}
\institute{Astronomical Institute, Utrecht University, PO Box 80000,
  3508 TA Utrecht, The Netherlands
  \and
  Stichting ASTRON, PO Box 2, 7990 AA Dwingeloo, The Netherlands
  \and
  Astronomical Institute `Anton Pannekoek',
  Kruislaan 403, 1098 SJ Amsterdam, The Netherlands  
  \and
  Physics Department, University of Vermont, Burlington, VT 05405,
  USA
}
\offprints{Joeri van Leeuwen\newline
  {\it Correspondence:} jleeuwen@astro.uu.nl\newline
  {\it More information:} http://www.astro.uu.nl/$\sim$jleeuwen/} 
\date{Received / Accepted}

\abstract{
  Both nulling and subpulse drifting are poorly understood
  phenomena. We probe their mechanisms by investigating how
  they interact in PSR~B0809+74. We find that the subpulse drift is not
  aliased but directly reflects the actual motion of the subbeams. The
  carousel-rotation time must then be over 200 seconds, which is much
  longer than theoretically predicted.
  \newline
  The drift pattern after nulls differs from the normal one, and using
  the absence of aliasing we determine the underlying changes in the
  subbeam-carousel geometry. We show that after nulls, the subbeam
  carousel is smaller, suggesting that we look deeper in the pulsar
  magnetosphere than we do normally. The many striking similarities
  with emission at higher frequencies, thought to be emitted lower
  too, confirm this. The emission-height change as well as the
  striking increase in carousel-rotation time can be explained by a
  post-null decrease in the polar gap height. This offers a glimpse
  of the circumstances needed to make the pulsar turn off so
  dramatically.


}

\maketitle

\section{Introduction}
In pulsars, the emission in individual pulses generally consists of
one or more peaks (`subpulses'), that are much narrower than the
average profile and the brightness, width, position and number of
these subpulses often vary from pulse to pulse.

\begin{figure}[t]
  \centering
  \includegraphics[]{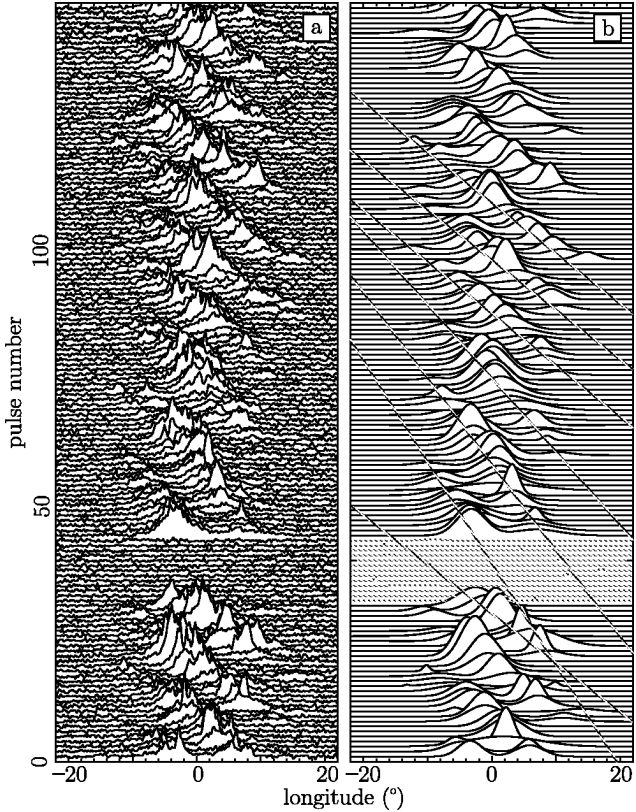}
  \caption{Observed and fitted pulse sequences. A
    window on the pulsar emission is shown for 150 
    pulses. One pulse period is
    $360^\circ$. The centre of the Gaussian that fits the pulse
    profile best is at $0^\circ$.
    {\bf a)} The observed pulse sequence, with a null after pulse 30.
    {\bf b)} The Gaussian curves that fitted the subpulses best. Nulls are
    shown in lightest gray, driftbands fitted to the subpulse pattern are
    medium gray.}
  \label{img:data.fit}
\end{figure}

In contrast, the subpulses in PSR~B0809+74 have remarkably steady widths
and heights and form a regular pattern (see
Fig. \ref{img:data.fit}a). They appear to drift through the pulse
window at a rate of $-0.09 P_2/P_1$, where $P_2$ is the average
longitudinal separation of two subpulses within one rotational period
$P_1$, which is $1.29$ seconds. Figure \ref{img:data.fit}a also shows
how the pulsar occasionally stops emitting, during a so-called null.

In this paper, we will interpret the drifting subpulse phenomenon in
the rotating carousel model \citep{rs75}. In this model, the pulsar
emission originates in discrete locations (`subbeams') positioned on a
circle around the magnetic pole. The circle rotates as a whole,
similar to a carousel, and is grazed by our line of sight. In between
successive pulses, the carousel rotation moves the subbeams through
this sight line, causing the subpulses to drift.

Generally, the average profiles of different pulsars evolve with
frequency in a similar manner: the profile is narrow at high
frequencies and broadens towards lower frequencies, occasionally
splitting into a two-peaked profile \citep{kis+98}. This is usually
interpreted in terms of `radius to frequency mapping', where the high
frequencies are emitted low in the pulsar magnetosphere. Lower
frequencies originate higher, and as the dipolar magnetic field
diverges the emission region grows, causing the average profile to
widen.

The profile evolution seen in PSR~B0809+74 is different. The movement
of the trailing edge broadens the profile as expected, but the leading
edge does the opposite. The profile as a whole decreases in width as
we go to lower frequencies until about 400 MHz. Towards even lower
frequencies the profile then broadens somewhat \citep{dls+84,
kis+98}. Our own recent observations of PSR~B0809+74, simultaneously
at 382, 1380 and 4880 MHz, confirm these results \citep{rrl+02}.
Why the leading part of the expected profile at 400 MHz is absent is
not clear. While \citet{bkk+81} suggest cyclotron absorption,
\citet{dls+84} conclude that the phenomenon is caused by a non-dipolar
field configuration. We will refer to this non-standard profile
evolution as `absorption', but none of the arguments we present in
this paper depends on the exact mechanism involved.

In a recent paper (\nocite{lkr+02}van Leeuwen et~al. 2002, henceforth
Paper I) we investigated the behaviour of the subpulse drift in
general, with special attention to the effect of nulls. We found that
after nulls the driftrate is less, the subpulses are wider but more
closely spaced, and the average pulse profile moves towards earlier
arrival. Occasionally this post-null drift pattern remains stable for
more than 150 seconds.

For a more complete introduction to previous work on PSR~B0809+74, as well
as for information on the observational parameters and the reduction
methods used, we refer the reader to Paper I. In this paper we will
investigate the processes that underly the post-null pattern
changes. We will quantify some of the timescales associated with the
rotating carousel model and map the post-null changes in the drift
pattern onto the emission region.

One of the interesting timescales is the time it takes one subbeam to
complete a rotation around the magnetic pole. This carousel-rotation
time is predicted to be of the order of several seconds in the
Ruderman \& Sutherland model. Only recently a carousel-rotation
time was first measured: \citet{dr99} find a periodicity
associated with a 41-second carousel-rotation time for PSR~B0943+10.

The second goal is to determine the changes in the emission region
that underly the different drift pattern we see after nulls. Mapping
this emission region could increase our insight into what
physically happens around nulls.

Achieving either goal requires solving the so-called aliasing problem:
as the subpulses are indistinguishable and as we observe their positions
only once every pulse period, we cannot determine their actual speed.

\section{Solving the aliasing problem}

The main obstacles in the aliasing problem are the under sampling of
the subpulse motion and our inability to distinguish between
subpulses. The pulsar rotation only permits an observation of
the subpulse positions once every pulse period. Following them through
subsequent pulses might still have led to a determination of
their real speed, but unfortunately the subpulses are so much alike
that a specific subpulse in one pulse cannot be identified in
the next, making it impossible to learn its real speed.

\begin{figure}[t]
  \centering
  \includegraphics[]{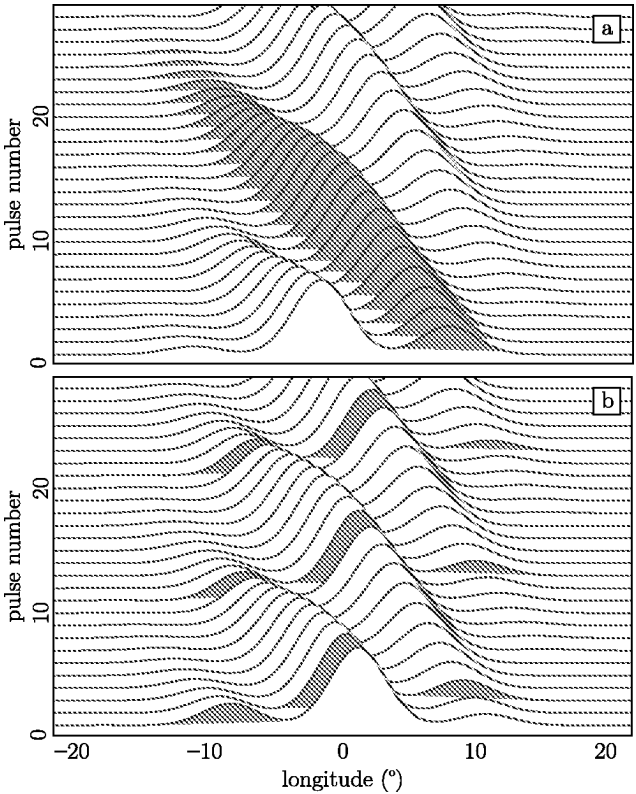}
  \caption{Different alias orders illustrated. We show two series of
  stacked simulated drifting subpulses. We have marked the
  subpulses formed by one particular subbeam with a darker colour. 
  {\bf a)} At a low subbeam speed, a single subbeam traces an entire
  driftband by itself. The driftrate is identical to the subbeam
  speed: alias order $0$.
  {\bf b)} At alias order $-1$, the subbeam speed is higher than the
  driftrate and in opposite direction.
}
  \label{img:alias_example}
\end{figure}

In Fig. \ref{img:alias_example} we show a simulation of subpulse
drifting, where we have marked all subpulses formed by a particular
subbeam with a darker colour. We use these simulations to discuss how
the driftrate, which is the observable motion of the subpulses through
subsequent pulses, is related to the subbeam speed, which cannot be
determined directly. In Fig. \ref{img:alias_example}a the speed of the
subbeams is low ($-0.09 P_2/P_1$) and identical to the driftrate. In
Fig. \ref{img:alias_example}b the subbeam speed is higher ($0.91
P_2/P_1$), but the driftrate is identical to the one seen in
Fig. \ref{img:alias_example}a. When the differences between subpulses
formed by various subbeams are smaller than the fluctuations in
subpulses from one single subbeam, these two patterns cannot be
distinguished from one another. In that case the subpulses within one
driftband, which seem to be formed by one subbeam, can actually be
formed by a different subbeam each pulse period (`aliasing').

To solve the aliasing problem for PSR~B0809+74 we follow driftrate changes
after nulls to determine the subbeam speed. Nulls last between 1 and
15 pulse periods, and in Paper I we have shown that for each null the
positions of the subpulses before and after the null are identical if
we correct for the shift of the pulse profile. So, as there is no
apparent shift in subpulse position, either the subbeams have not
moved at all, or their movement caused the new subpulses to appear
exactly at the positions of the old ones.

As the lengths of the nulls are drawn from a continuous sample it is
highly unlikely that the subbeam displacement is always an exact
multiple of the subpulse separation: only a total stop of the subbeam
carousel can explain why the subpulse positions are always unchanged
over the null. At some point after the null, however, the subbeams
have accelerated, and the drift pattern has returned to normal. In
Fig. \ref{img:data.fit} we see how, after a null, the driftrate
increases to its normal value in about 50 pulses.

There are two scenarios for this subbeam acceleration. The first we
will call gradual speedup.  Here the changes in the subbeam speed
occur on timescales larger than $P_1$. The second we will call
instantaneous, as the entire acceleration happens within $1P_1$,
effectively out of sight.

\begin{figure*}[t]
  \centering
  \includegraphics[]{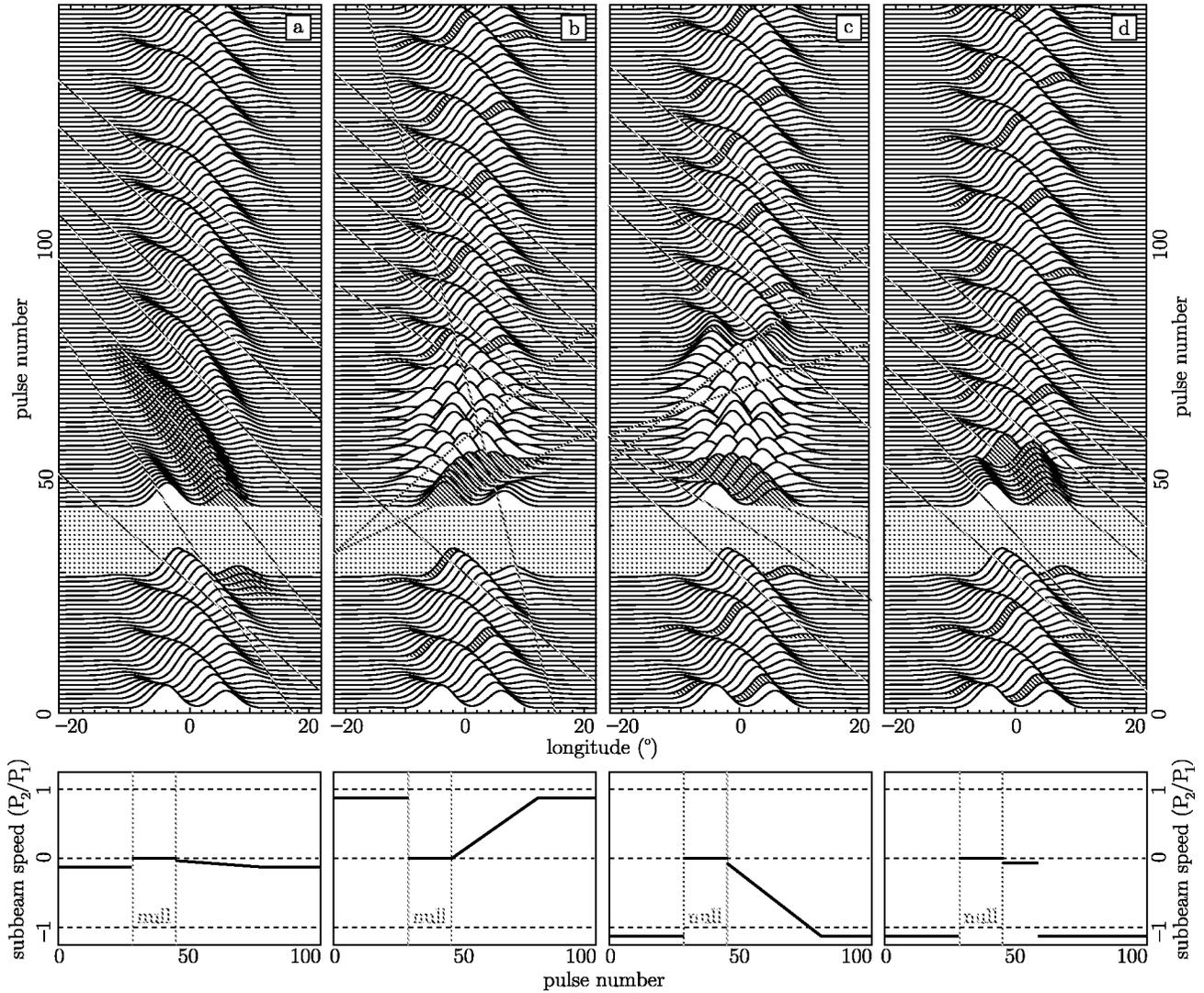}
  \caption{Simulated scenarios for subbeam speedup. The top panels show
    the resulting drift patterns.  Nulls are shown in lightest gray,
    driftbands fitted to the subpulse pattern are overdrawn in medium
    gray. We have marked the subpulses formed by one particular
    subbeam with a darker colour.  The bottom panels show the subbeam
    speed versus the pulse number.
    {\bf a)} Gradual acceleration to a subbeam speed equal to the
    driftrate (alias order 0). 
    {\bf b)} Gradual acceleration to a subbeam speed larger than the
    driftrate and in opposite direction (alias order -1)
    {\bf c)} Gradual acceleration to a subbeam speed larger than the
    driftrate and in the same direction (alias order 1)
    {\bf d)} Instantaneous acceleration, again to alias order 1.
}
  \label{img:sim.speedup}
\end{figure*}

In Fig. \ref{img:sim.speedup} we show four simulated pulse sequences
with different speedup parameters. In all cases, we simulate a drift
pattern like that of PSR~B0809+74. During a null, from pulses 30 to 45,
there is no subbeam displacement. Immediately after the null the
subbeams build up speed, and each pulse period we translate the
subbeam displacement to a change in subpulse position. Although the
final driftrate is the same for all scenarios ($-0.09 P_2/P_1$), the
subbeam speeds differ considerably. The bottom four graphs show these
speeds for each scenario. In the top four diagrams we have marked the
subpulses from one subbeam with a darker colour for clarification.

Let us look at the case of gradual speedup to alias order $0$,
where the subbeam speed is the same as the driftrate
(Fig. \ref{img:sim.speedup}a). In this case the driftrate will
gradually increase and form a regular driftband pattern, much like the
pattern found in the observations.

Next, we investigate a subbeam acceleration to a slightly higher
speed. At $0.91 P_2/P_1$, Fig. \ref{img:sim.speedup}b shows alias mode
$-1$, which is the simplest configuration in which the subbeams move
opposite to the subpulse drift. Right after the null the drift
consequently commences in this opposite direction. When the subbeam
speed nears the first aliasing boundary $0.5 P_2/P_1$, the subpulses
seem to move erratically through the window. After the subbeam speed
passes the first alias boundary, the subpulse drift resumes its normal
direction, and as the subbeam speed approaches $0.91 P_2/P_1$, the
drift pattern returns to normal.

Also in the `alias order 1' scenario (Fig. \ref{img:sim.speedup}c),
which is the simplest aliased mode in which the subbeams move in the
same direction as the subpulses, the subpulses wander when the subbeam
speed nears an alias boundary. Such a disturbance of the drift pattern
turns out to be present in all simulations of non-zero alias
orders. Because the observed pulse sequences always show smooth,
non-wandering driftbands (like in Fig. \ref{img:data.fit}a), we
conclude that the subbeams cannot accelerate gradually to a high
speed.

With instantaneous acceleration the subbeam speed switches
suddenly. Most likely it will do so when the pulsar beam faces away
from us. After we see pulse number $n$, the subbeams will move slowly
for a certain time, quickly speed up, and then move fast until we see
the subpulses of pulse $n+1$ appear. The actual speedup can occur any
moment between seeing pulses $n$ and $n+1$. This means that the
displacement of the subbeams can vary from very little (speedup just
before pulse $n+1$) to a lot (speedup right after pulse $n$). The
accompanying changes in subpulse position will then be evenly
distributed between 0 and $P_2$; the subpulse positions after the
speedup will not be related to those before.

If the subbeams accelerate instantaneously at the end of a null,
before the first pulse can be observed, the subpulse phases are not
preserved over the null. As this is opposite to what is observed, this
possibility is ruled out.

If the subbeam acceleration occurs a few pulses after a null, we
expect that the change in driftrate will nearly always be accompanied
by a sudden change in the longitude of the driftband, as illustrated
in Fig. \ref{img:sim.speedup}d for a final subbeam speed of $-1.09
P_2/P_1$. The absence of a significant number of such sudden shifts in
the data indicates that the subbeam carousel of PSR~B0809+74 does not speed
up instantaneously to a high speed.

As only the gradual speedup to alias order 0 can explain the observed
drift patterns, the drift seen in the subpulses of PSR~B0809+74 directly
reflects the movement of the subbeams, without any aliasing.

\section{Discussion}
\subsection{Alias order of other pulsars}
The subbeam speedup in PSR~B0809+74 follows the simplest scenario possible:
it is gradual, and the subpulse drift is not aliased. In other pulsars
this might not be the case, and for those we predict drift-direction
reversals or jumps in driftband longitude during the subpulse-drift
speedup phase after nulls.

\subsection{Subbeam-carousel rotation time}
Having resolved the aliasing of the subpulses' representation of the
subbeams, we know that the subbeams move at $-0.09 P_2/P_1$. Such a
low rotation rate means it will take each subbeam $11P_1$ to move to
the current position of its neighbour. The following estimate of
the number of subbeams then leads directly to the carousel-rotation
time.

Towards the edges of the profile the sight line and the carousel move
away from each other, and a subbeam will cease to be visible when the
sight line no longer crosses it. When the subbeams are small, or widely
spaced compared to the curvature of the carousel, the sight line will
cross only few. In that case there will not be many subpulses visible
in one pulse. If, however, the subbeams are large and closely spaced
compared to the carousel curvature, the sight line passes over more
subbeams in one traverse, leading to many subpulses per pulse.

In most of the pulses of PSR~B0809+74 we observe two subpulses,
occasionally we discern three. By combining this with the ratio of the
subpulse width and separation we find there must be more than 15
subbeams on the carousel. As one subbeam reaches the position of its
neighbour in $11P_1$, a 15-subbeam carousel rotates in over 200
seconds.

Persistent differences in the properties of individual subbeams should
introduce a long term periodicity in the pulse sequences, at the
carousel-rotation frequency. Thus far, no such periodicity has been
found, which is not surprising as the periodicity can only be measured
if the lifetime of the subbeam characteristics is longer than the
carousel-rotation time.

In PSR~B0943+10, the only pulsar for which we know how the subbeams vary
\citep{dr99,dr01}, the associated timescales are in the order of 100
seconds. If we assume their lifetimes are comparably long in PSR~B0809+74,
the subbeams will have lost most of their recognisable traits once
they return into view after one carousel rotation. Having regular
pulse sequences that contain several tens of rotation times might
still show some periodicity at the carousel-rotation
frequency. Unfortunately nulls have a destructive influence on the
driftband pattern and possibly on the subpulse characteristics. This
has thus far made it impossible to observe bright sequences longer
than 3 times our lower limit on the rotation time of 200 seconds.

After the prediction that the rotation time in most pulsars would be
on the order of several seconds \citep{rs75}, the observation that it
is 41 seconds in PSR~B0943+10 was somewhat surprising. The suggested
dependence of the carousel-rotation time on the magnetic field
strength and the pulse period predicts that the rotation time in
PSR~B0809+74 should be roughly 4 times smaller. The circulation time we
observe, several hundreds of seconds, shows that the theoretical
predictions are incorrect in absolute numbers as well as in the
scaling relations they propose.

\begin{figure}
  \centering
  \includegraphics[]{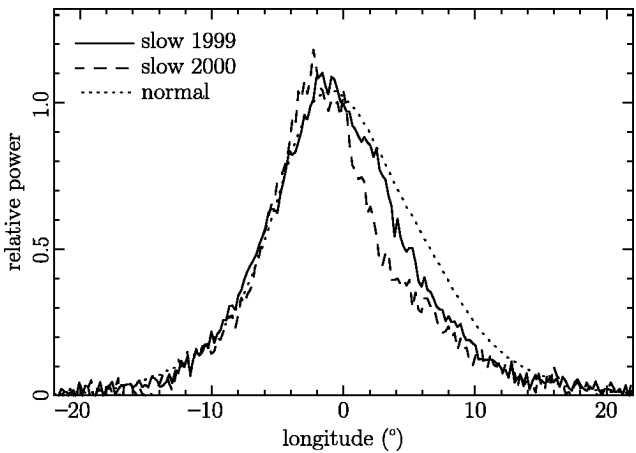}
  \caption{Pulse profiles for the normal mode and the two longest slow
  drifting sequences after nulls described in Paper I. The broadening
  of the subpulses causes the slow-drifting profile to outshine the normal
  one. Here all three profiles have been scaled to the same height to
  show the change in the pulse location and shape.}
  \label{img:prof}
\end{figure}

\subsection{Emission region geometry}
In Paper I we found many interesting changes in the driftband pattern
after nulls. The subpulses reappear $5-10\%$ broader and $15\%$ closer
together, which results in a brighter average profile. The individual
subpulses move $1.5^{\circ}$ towards earlier arrival, as does their
envelope, the average profile, with $1.1^{\circ}$
(Fig. \ref{img:prof}). The subpulses drift $50\%$ more slowly than
usual. Occasionally, this slow drifting pattern remained stable for
over 100 pulses.

\begin{figure*}[t]
  \centering
  \includegraphics[]{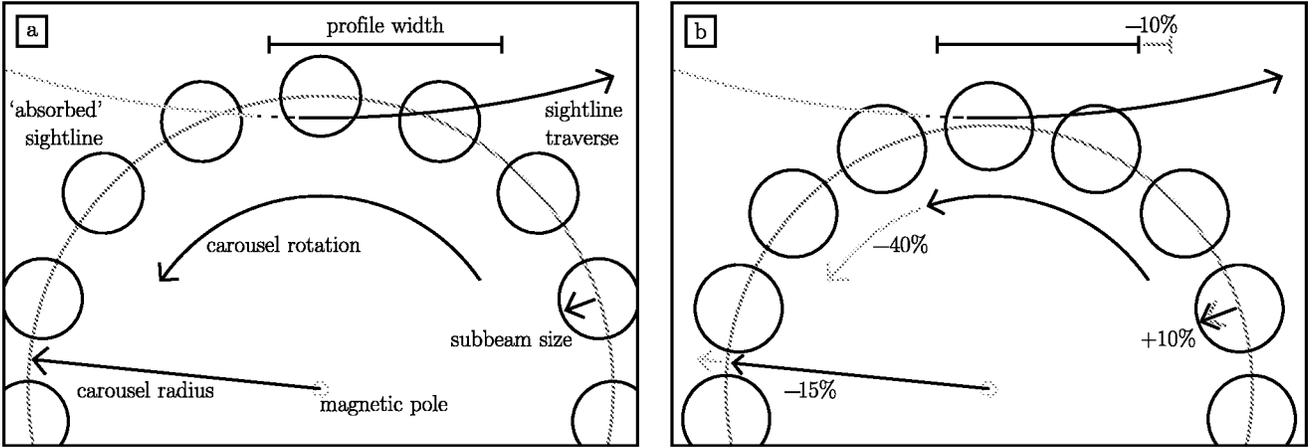}
  \caption{Subbeam carousel geometry in the normal and the post-null
  configuration.  {\bf a)} The normal configuration, with a sight line
  that is partially obscured by the absorption.  {\bf b)} After nulls
  the carousel is smaller. As their number is unchanged, the subpulses
  have moved closer together. On the non-absorbed, trailing edge, the
  pulse window moves inward. The leading edge of the profile is
  located where the absorption ends and does not move. The subpulses
  are wider and revolve less fast.}

  \label{img:carousel}
\end{figure*}

Knowing, as we do now, that there is no aliasing involved, we can
identify single subpulses with individual subbeams. This implies that
we can interpret the drift pattern changes in terms of the
subbeam-carousel geometry (Fig. \ref{img:carousel}). We will do so for
each of the altered characteristics.

The first thing we note is that, relative to the shift of the average
profile, the positions of the subpulses are unchanged over nulls. This
means that the subbeam carousel has not rotated during the null.

After nulls, the subpulses are positioned about 15\% closer
together. In principle, this could be the result from a change in
subbeam speed. At a certain moment, a particular subbeam will be
pointing towards the observer. It takes some time before the pulsar
has rotated the next subbeam into the observer's view, and during this
time the subbeams themselves have moved, too. This motion translates
directly to a change in the longitudinal subpulse separation. Yet,
as we have shown that subbeam speed is low, this effect is negligible.

A more significant change in $P_2$ could be caused by an increase in
the number of subbeams on a carousel of unchanged size. The other
option is that a decrease in the carousel radius moves the subbeams
closer together.

In the first scenario the number of subbeams would have to change
during the null, causing the subbeam placement to change considerably.
This would lead to subpulse-position jumps over the null. In that
case, we would not expect the phases of the subpulses to be as
unchanged over the nulls as is observed. Secondly, some time after the
null the new configuration would have to return to normal. This means
subbeams would have to appear or disappear. We see no evidence of this
in any of the 200 nulls we observed. The change in the subbeam
separation $P_2$ can therefore not be due to a changed number of
subbeams.

In the second scenario the carousel radius decreases by 15\%, but the
number of subbeams remains the same.  As the subbeams now share a reduced
circumference, their separation also decreases.

The contraction of a carousel causes all subpulses to move towards the
longitude of the magnetic axis. If we combine this contraction with
the aforementioned absorption, we can immediately explain the observed
shift to earlier arrival of the subpulses and their envelope, the
average profile: with only the trailing part unabsorbed, all visible
subpulses move towards earlier arrival (Figs. \ref{img:prof} and
\ref{img:carousel}).

In general, the subpulses farthest from the longitude of the magnetic
axis will move most, while those at it, if visible, will remain at
their original positions. Enlargement of our sample of subpulse
positions can therefore indicate were the magnetic axis of PSR~B0809+74 is
located. This would immediately indicate how much of the pulse profile
is `absorbed' and illuminate the thus far much debated alignment of the
pulse profiles at different frequencies.

Normally, the edges of a pulse profile indicate where the overlap of
the sight line and the subbeam carousel begins and ends. The
contraction of a carousel will then move both edges inward. Yet when
the first part of the profile is absorbed, the leading edge reflects
the end of the absorption. If this edge is near the middle of the
sight-line traverse over the carousel, it will not be affected by a
reduction of the carousel size. In that case we will see a change in
the position of the trailing edge but not in the leading one, which
is exactly what we observe in PSR~B0809+74 (Fig. \ref{img:prof}).

The wider subpulses we see after a null translate directly to
wider subbeams. We do note that a change in carousel radius leads to a
new sight-line path, which may have an impact as well.

The significant chance that a null starts or ends within the pulse
window of the pulses that surround it, would on average make these
pulses dimmer than normal ones \citep{la83}. Quite unexpectedly
however, the pulses after nulls were found to be brighter than
average. This is easily explained in our model: the change in the
carousel geometry causes the subpulses to broaden and become more
closely spaced, while their peak brightness remains the same. This
leads to the post-null pulse-intensity increase that is observed.

\subsection{Post-null polar gap height}
If we follow the radius-to-frequency-mapping argument discussed in the
introduction, then the reduction of the carousel size implies that the
emission after a null comes from lower in the pulsar magnetosphere
than it does normally. The reduction of the subbeam separation
supports this idea, but the broadening of the subpulses seems
inconsistent. 

At this stage a comparison with the pulse profiles and drift patterns
observed at higher frequencies seems promising. Emission at higher
frequencies is also supposed to originate lower in the pulsar
magnetosphere, so similar effects may point at one cause: a decrease
in emission height.

Much like the post-null profile, the average profile at higher
frequencies moves towards earlier arrival \citep{rrl+02}. Comparing
drift patterns, we see that at higher frequencies the driftrate
decreases and that the subpulses broaden and move closer to one
another \citep{dls+84}, strikingly similar to the behaviour we find
after nulls.

Only one anomaly remains: when comparing drift patterns at
different frequencies, the changes in driftrate and subpulse
separation $P_2$ always counterbalance, leaving the carousel rotation
time $\hat{P}_3$ unchanged. This means that the subbeams we observe at
a certain frequency are cuts through rods of emission, intersections
at the emission height associated with that frequency. The rods rotate
rigidly but diverge with increasing height. At each height $\hat{P}_3$
is the same, while the speed and separation of the subbeams do
change.

In contrast to this normal invariance of $\hat{P}_3$, we see it change
considerably after nulls, an increase that cannot be explained by a
change in viewing depth. We therefore expect that the disturbance that
caused the change in this depth, can also account for the 50\%
increase in $\hat{P}_3$.

Usually, both the subbeam separation $P_2$ and the emission height are
assumed to scale with the polar gap height $h$ \citep{rs75,mgp00}. The
15\% decrease in $P_2$ would thus be due to a identical fractional
decrease in the gap height, which should also cause an equal decrease
in emission height.

If this change in gap height could also account for the observed
increase of the carousel rotation time $\hat{P}_3$, all post-null
drift-pattern changes can be attributed to one single cause.

$\hat{P}_3$ is thought to scale as $h^{-2}$. For the inferred 15\% gap
height decrease this predicts a 40\% increase in $\hat{P}_3$, nicely
similar to the 50\% we find.

With one single cause we can therefore explain both the puzzling
well-known phenomena (the driftrate decrease after nulls, the bright
first active pulse) and the newly discovered subtle ones (the change
in the position of the average profile, the decrease of the subpulse
separation and the subpulse-width increase). This post-null decrease
in gap height offers a glimpse of the circumstances needed to make the
pulsar turn off so dramatically.

\section{Conclusions}

We have shown that the drift of the subpulses directly reflects the actual
motion of the subbeams, without any aliasing.

In other pulsars with drifting subpulses this may be different: for
those we predict drift-direction reversals or longitude jumps in the
post-null drift pattern.

We find that the carousel-rotation time for PSR~B0809+74 must be long,
probably over 200 seconds. The expected lifetime of the subbeam
characteristics is less, which explains why thus far no periodicity
from the carousel rotation could be found in the pulse sequence.

The rotation time we find is larger than theoretically predicted, not
only in absolute numbers but also after extrapolating the rotation
time found in PSR~B0943+10. Both the magnitude and the scaling relations
that link the carousel-rotation time to the magnetic field and period
of the pulsar are therefore incorrect.

When the emission restarts after a null the drift pattern is
different, and having determined the alias mode, we identify the
underlying changes in the geometry of the subbeam carousel. A
combination of a decrease in carousel size and `absorption' already
explains many of the changes seen in the post-null drift pattern.

The resemblance between the drift pattern after nulls and that seen at
higher frequencies, thought to originate at a lower height, is
striking. Assuming that similar effects have identical causes leads us
to conclude that after nulls we look deeper in the pulsar
magnetosphere, too.

Both this decrease in viewing depth and the striking increase in the
carousel rotation time can be quantitatively explained by a post-null
decrease in gap height.

\begin{acknowledgements}
We thank Marco Kouwenhoven for his help with the observations and
Frank Verbunt, Russell Edwards and Patrick Weltevrede for constructive
discussions.
\end{acknowledgements}

\end{document}